\documentclass{article}

\usepackage{indentfirst}
\usepackage{amsmath}
\usepackage{graphicx}
\usepackage[
  left=1in,
  right=1in,
]{geometry}

\title{A new thermodynamic language for colloid systems}
\author{Junfeng Zhou\thanks{1155187531@link.cuhk.edu.hk}, Geliang Zhu \& Lei Xu\thanks{xuleixu@cuhk.edu.hk}}
\date{\today}

\begin{document}
	\maketitle
	
	\begin{abstract}
		A simple framework is presented for unified applications in various fields of colloidal research, with minimal additional concepts \& definitions. Several case studies concerning glass transition \& crystallization are provided under the minimalist version, upon which adaptations can be made to suit more complicated topics. Major factors influencing accuracy are also discussed.
	\end{abstract}
	\section{Introduction}
	Colloid studies have a long-standing problem: overflow of problem-specific definitions. Starting from coordinates as initial inputs, advanced variables are calculated in various ways. They often differ from one topic to other (crystallization \& glass transition \& grain boundary \& amorphous intermediates etc.). Even same-topic parameter calculations sometimes diverge materially. This leads to both lack of complete inter-field pictures and difficulty of intra-field comparisons. From a theoretical point of view, this is certainly unsatisfactory.
	
	The reason is simple: Although colloid particles serve well as magnified $\mu m$-scale models for atoms\cite{Poon2004}, their governing theoretic framework is not a direct copy of atomic counterparts. Thermodynamic variables are not originally designed for mesoscopic systems. Concept like entropy does explain the driving force of hard-sphere colloidal crystallization. Critical volume fractions can also be calculated analytically\cite{Onsager1949}\cite{Dinsmore1995}. However, easy application of traditional thermodynamics ends here. Descriptive \& predictive power of classic macroscopic thermodynamic functions quickly diminishes due to both insufficient data caused by mesoscopic systems and excessively short-range nature of colloidal attractive \& repulsive interactions. Another route, which is the $N$-particle stochastic diffusive process formalism based on the Smoluchowski equation for joint probability distribution $P\left(\left\{\vec{r}\right\}, t\right)$, also proved difficult. If curved surface is involved, a whole set of covariant geometry tools like $\Delta$=$\left(1/\sqrt{g}\right)\partial_{a}\sqrt{g}g^{ab}\partial_{b}$ is introduced\cite{SolanoCabrera2025Colloids}. Forced by reality, researchers resort to innovative definitions \& data processing techniques, leading to the problem we mentioned\cite{Larsen2020}. Representative examples include common neighbor analysis (CNA)\cite{BlaistenBarojas1984}\cite{UrrutiaBanuelos2016}\cite{Deng2018}, centrosymmetry parameter (CSP)\cite{Kelchner1998} and coordinate polyhedron method (CPM)\cite{Du2017}. AI-based inventions also begin to emerge recently.
	
	Here in this work, a unified theoretical framework is presented, along with its applications in Three different problems. $(A)$ structural origin of glass transitions, $(B)$ supercooling and $(C)$ curvature. The framework is designed to be colloid-native with minimal additional artificial definitions \& complications, while at the same time compatible with traditional thermodynamics to facilitate inter-field applications.
	
	\section{Unified Framework For Colloids}
	Framework construction requires a first principle, which we believe is energy. Traditional thermodynamic system has two fundamental system-environment interaction schemes: heat exchange and work. Correspondingly, two intensive-extensive variable pairs are set, which is $T-S$ and $P-V$. 
	
	In colloid systems, heat exchange still exists because of Brownian motions. Its corresponding variable pair is $f\left(\phi\right)-S_{2}$. Temperature is conventionally substituted by supercooling $f\left(\phi\right)$, which is a function of volume fraction $\phi$. Two-body excess entropy $S_{2}$ substitutes the traditional one because it's coordinate-based and colloid-native\cite{NettletonGreen1958} (higher order terms exist, but two-body term contributes $85\%-90\%$).
	
	Colloid systems, as particle suspensions sealed in tubes, no longer perform pressure-volume "work" in a physically meaningful way. Instead the other system-environment interaction scheme is the aggregate effect comprised of each individual particle-solvent hydrodynamic interaction (HI). System-wise, this scheme is modulated by rearrangements. Particles rearrange towards minimal-energy configurations, during which stabilization occurs \& kinetic energy is dissipated \& HI weakens. Since this rearrangement process is always signaled by order. We select the most widely used order parameter to construct the variable pair. That is bond orientation order parameter $Q_{6}$ (BOO)\cite{SteinhardtNelsonRonchetti1983}, and the resulting variable pair becomes $Q_{6}-\varphi$. Using $Q_{6}$ is not mandatory. Other dimensionless order parameters can be used freely. $\varphi$ is newly-defined and has no clear counterpart in traditional thermodynamics. Its physical meaning is of primary importance and will be explored below. (In equation \eqref{eq:3})
	
	To complete the framework, particle exchange is considered as $\mu\left(\phi\right)-N$. Here chemical potential $\mu$ is also dependent on volume fraction $\phi$ since most particle additions / deletions are driven by diffusion. Put together, we arrive at the fundamental equation:
	\begin{equation}
		\label{eq:1}
		dU=f\left(\phi\right)dS_{2}-Q_{6}d\varphi+\mu\left(\phi\right)dN
	\end{equation}
	
	Along with the following continuity equations:
	\begin{equation}
		\label{eq:2}
		\begin{aligned}
		    \frac{\partial U}{\partial t} & = -\nabla\cdot\vec{J_{U}} \\
		    \frac{\partial S_{2}}{\partial t} & = -\nabla\cdot\vec{J_{S_{2}}} + \frac{\partial S_{2}^{c}}{\partial t} \\
		    \frac{\partial N}{\partial t} & = -\nabla\cdot\vec{J_{N}} \\
		\end{aligned}
	\end{equation}
	
	To express colloid system's potential for non pressure-volume work, which is predominantly HI modulated by rearrangements, we define the "Gibbs free energy" as:
	\begin{equation}
		\label{eq:3}
		\begin{aligned}
			G & = U -f\left(\phi\right)S_{2}+Q_{6}\varphi = \mu N \\
			dG & = -S_{2}df\left(\phi\right)+\varphi dQ_{6}+\mu\left(\phi\right)dN \\
			\varphi & = \frac{\partial G}{\partial Q_{6}}
		\end{aligned}
	\end{equation}
	
	Thus the physical meaning of $\varphi$ is "Gibbs free energy" change rates with respect to order measured by $Q_{6}$. It's an extensive variable because larger systems absorb \& release more energy per unit order change. Intuitively, this can be rephrased as depletion / absorption rates in the system's potential for more rearrangements as current ones take place. More generally, $\varphi_{\alpha} = \partial G/\partial\alpha$ if $\alpha$ is chosen to measure order.
	
	Based on this framework \eqref{eq:1}\eqref{eq:2}, our first task is to express the newly-defined \& somewhat abstract $\varphi$ in usual \& calculable variables. This is exactly the idea behind Gibbs-Duhem equation. Thus, following the same derivation, we arrive at its counterpart:
	\begin{equation}
		\label{eq:4}
		\varphi=\frac{Nd\mu\left(\phi\right)+S_{2}df\left(\phi\right)}{dQ_{6}}
	\end{equation}
	
	Next, when assuming $\partial N/\partial t=0$ and $\partial\varphi/\partial t = -\nabla\cdot\vec{J_{\varphi}}$, simple derivation leads to a useful expression for entropy sources.
	\begin{equation}
		\label{eq:5}
		\frac{\partial S_{2}^{c}}{\partial t}=-\frac{Q_{6}}{f\left(\phi\right)}\nabla\left[\frac{f\left(\phi\right)}{Q_{6}}\right]\cdot\vec{J_{S_{2}}}+\frac{Q_{6}}{f\left(\phi\right)}\nabla\frac{1}{Q_{6}}\cdot\vec{J_{U}}
	\end{equation}
	
	Of course, eliminating sources \& sinks of $\varphi$ could be problematic. If so, equation \eqref{eq:5} will render unreasonable results.
	
	\section{Energy Gap}
	Here we present a universally applicable procedure in which simple reasoning enables us to analyze energy gaps. With coordinates \& inter-particle interactions known, dynamical matrix \& eigenmode $\left\{d\vec{r}\right\}_{i}$ can be calculated. Then an ensemble of imaginary excitations can be formed as eigenmode superposition $\sum A_{i}\left\{d\vec{r}\right\}_{i}$. $\left\{d\vec{r}\right\}_{i}$ has an excitation possibility of $min\left[1, exp\left(-dE_{i}/\alpha\right)\right]$. If excited, $A_{i}$ is randomly chosen in $\left[0, D\right]$, in which $D$ is particle diameter.
	
	Based on these imaginary excitations, $dQ_{6}$ is known and $\varphi$ can be calculated from equation \eqref{eq:4}. We focus on the \textit{minimum value} of $\varphi dQ_{6}$ as $A\left(\varphi\right)$. According to our definition of "Gibbs free energy" or equation \eqref{eq:3}, $A\left(\varphi\right)$ gives maximum loss of energy achievable through rearrangements (Assuming negativity, which is satisfied in examples below). Since imaginary excitations $\sum A_{i}\left\{d\vec{r}\right\}_{i}$ are unlimited, maximum energy loss accurately depicts energy surplus of current configuration compared with the global minimal-energy configuration. The latter is fixed. It can serve as a reference point. Thus we can measure energy gap as:
	\begin{equation}
		\label{eq:6}
		dE_{\alpha\rightarrow\beta}\left(\varphi\right) = A_{\alpha}\left(\varphi\right) - A_{\beta}\left(\varphi\right)
	\end{equation}
	
	The above calculation scheme can also be performed on real recorded excitations (experiments \& simulations), which are limited but more convincing. In both cases, accuracy of equation \eqref{eq:6} depends on attainability of the global minimal-energy configuration. Thus, for systems in or near equilibrium this calculation is generally more accurate. Early stage of crystallization \& glass formation remains challenging, but equation \eqref{eq:6} serves as a valuable reference nonetheless. Also it's possible that error cancellation makes $dE$ more accurate than $A_{\alpha}$. This scheme eliminates the need for neighborship definitions \& $\Delta\sum V\left(r\right)$ completely. Finally, it must be noted that only experiment data incorporate HI perfectly. In simulation \& imaginary excitations it's either approximated or non-existent completely.
	
	\begin{figure}[htbp]
		\centering
		\includegraphics[width=1\linewidth]{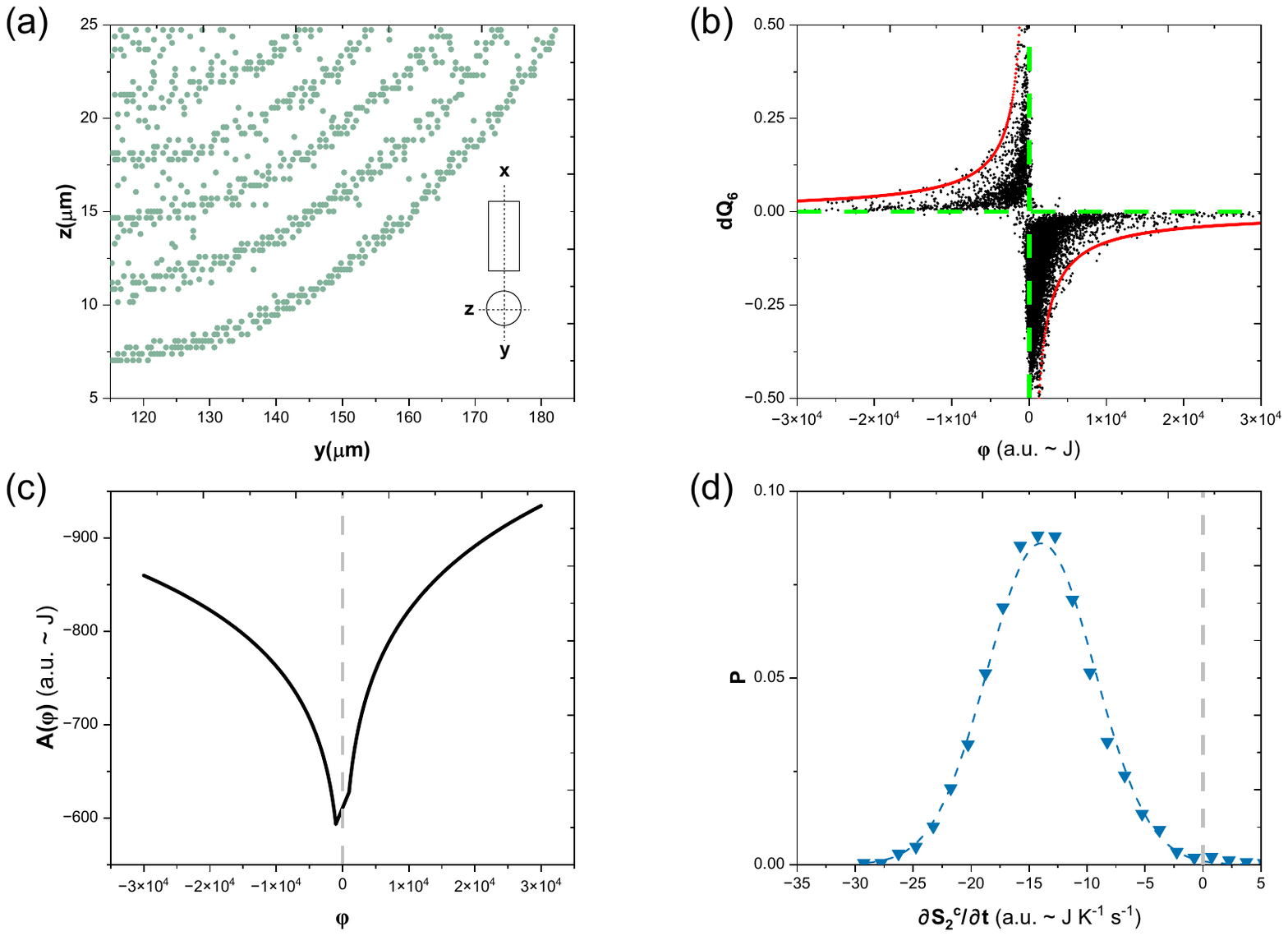}
		\caption{\textbf{Demo calculations.} Representative results on a mature Yukawa colloidal crystal sealed in a circular tube. \textbf{(a) sample.} Field of view contains one quarter of the circular tube. Only $y-z$ cross-section is shown for clarity. Axial direction $x$ is omitted. \textbf{(b) $\mathbf{\varphi-dQ_{6}}$ pairs.} Distribution of generated $dQ_{6}$ and calculated $\varphi$ using equation \eqref{eq:4} under imaginary excitations $\sum A_{i}\left\{d\vec{r}\right\}_{i}$. Most pairs lead to decrease in "Gibbs free energy", as shown in equation \eqref{eq:3}. An envelope appears corresponding to maximum energy loss achievable through rearrangements at various values of $\varphi$. \textbf{(c) $\mathbf{A\left(\varphi\right)}$.} Energy surplus to the global minimum (indicated by maximum energy loss), from the envelope in (b). Vertical coordinate is reversed since larger negativity means bigger energy surplus compared with the global minimum, or $A\left(\varphi\right) = -dE$. $dE$ is smallest when $\varphi=\partial G/\partial Q_{6}\rightarrow0$, which is a clear signal of approaching minimum. \textbf{(d) aggregate spontaneous entropy generation rate $\mathbf{\sum\partial S_{2}^{c}/\partial t}$.} Calculated by equation \eqref{eq:5} under imaginary excitations $\sum A_{i}\left\{d\vec{r}\right\}_{i}$, result is a normal distribution mostly in negative regime. This signals stability, which is reassuring for a mature sample in equilibrium. This demo is calculated with $f\left(\phi\right)=10\sqrt[3]{1/2\phi}$, $\mu$$\left(\phi\right)=10\phi$ and $\alpha=2.5$.}
		\label{fig:figure1-demo}
	\end{figure}
	
	For demonstrations, a Yukawa colloidal crystal in equilibrium is sealed in circular tube. It's scanned under Leica SP8 confocal microscopy. The above scheme is applied to produce $\varphi-dQ_{6}$ pair distributions and $A\left(\varphi\right)$. No morphology transition is studied since this crystal is already in deep equilibrium. Thus $A\left(\varphi\right)$ here is system-wise for all particles in field of view (FOV). Also spontaneous entropy generation $\partial S_{2}^{c}/\partial t$ is calculated using equation \eqref{eq:5}. For simplicity, we examine the probability distribution of global sum $\sum\partial S_{2}^{c}/\partial t$. Both calculations are based on imaginary excitations $\sum A_{i}\left\{d\vec{r}\right\}_{i}$.
	
	As seen in figure \ref{fig:figure1-demo}, calculation results are reassuring \& reasonable. $\varphi-dQ_{6}$ pairs almost lie entirely in the $\varphi dQ_{6} < 0$ regime (Figure\ref{fig:figure1-demo}.b), which means $dG<0$ according to equation \eqref{eq:3}. At the same time, system-wise spontaneous entropy generation is mostly negative (Figure\ref{fig:figure1-demo}.d). These two facts are reassuring that imaginary excitations $\sum A_{i}\left\{d\vec{r}\right\}_{i}$ are realistic, under which the sample crystal in equilibrium will self-stabilize rather than self-destroy its symmetry \& order. Maximum energy loss or energy surplus $A\left(\phi\right)$ is shown as an envelope in Figure\ref{fig:figure1-demo}.b and demonstrated in Figure\ref{fig:figure1-demo}.c. The surplus is smallest when $\varphi=\partial G/\partial Q_{6}\rightarrow0$. This is reasonable since $\partial G/\partial Q_{6}=0$ is a good indicator of minimal-energy configuration and should be accompanied by low energy surplus. Ideally, we expect $A\left(0\right)=0$ if calculations are absolutely accurate in terms of exploring the potential energy landscape (PEL).
	
	\section{Case Study}
	\subsection{TEAG \& Glass transition}
	Structural origin of glass formation is interesting\cite{Xia2015}. What mechanism prevents crystallization and causes dynamical arrest? Mode coupling theory (MCT) offers a signal, the non-ergodicity parameter, but it's based on system-wise intermediate scattering function (ISF). What happens at particle-level remains elusive. Many attempts to characterize cage-breaking behavior have been made\cite{Weeks2000Science}\cite{Weeks2002}\cite{Noji2025}, but they also suffer from excessive problem-specific definitions.
	
\begin{figure}[htbp]
		\centering
		\includegraphics[width=1\linewidth]{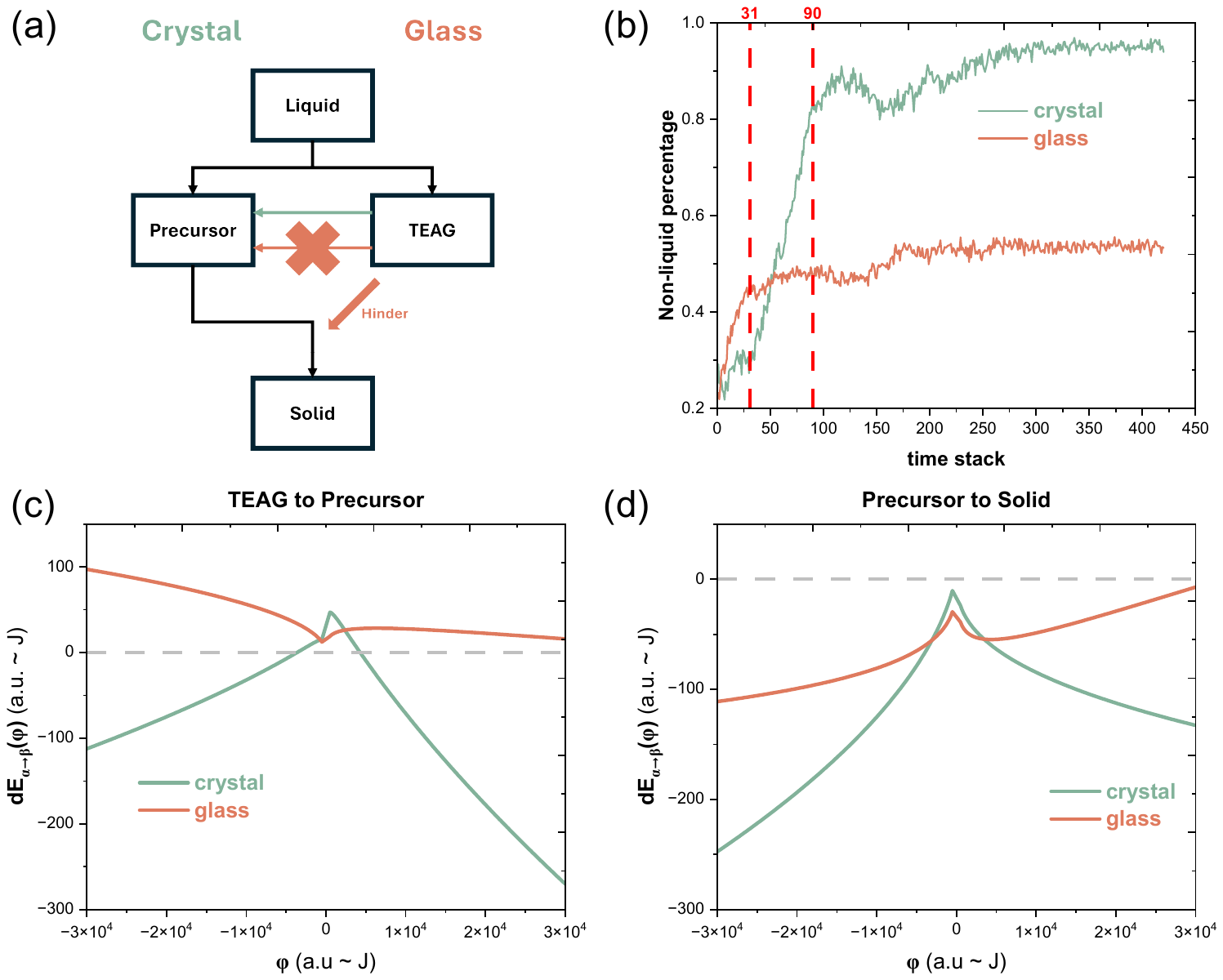}
		\caption{\textbf{Case study: glass transition.} Framework applications in explaining structural origins of glass. \textbf{(a) TEAG-based paradigm.} Tetrahedron-aggregate (TEAG) serves as the key in glassy dynamics. If TEAG can not readily transform into precursors \& hinders subsequent precursor-solid transitions, glass will form. \textbf{(b) critical periods.} Portion of non-liquid particles in two simulated LJ samples, one forming a glass and the other a crystal. Significant difference emerges during stack $31-90$, which is called critical period and on which calculations are based. \textbf{(c) TEAG-precursor transitions.} A clear energy barrier appears in glass but not in crystal. \textbf{(d) Precursor-solid transitions.} This transition is also less favorable in glass, presumably due to the presence of untransformed TEAG. In all cases, $dE\rightarrow0$ when $\varphi\rightarrow0$.}
		\label{fig:figure2-glass}
	\end{figure}

	Here we follow a recently established paradigm, and show our framework can be utilized to prove it without introducing any new definitions. According to this explanation, the key to glass formation is a kind of local structure called tetrahedron-aggregate\cite{Tsurusawa2023}\cite{LonialWeeks2026}, or TEAG (Figure\ref{fig:figure2-glass}.a). In a simulated LJ glass sample, TEAG has difficulties in transforming into precursors and also hinders subsequent precursor-solid transitions. On the contrary, in simulated LJ crystal sample, TEAG readily transforms into precursors and then solids. The process is completed swiftly. Macroscopic fates of two systems seem to differ because of microscopic fates of TEAG particles. Thus, we calculate $dE_{\alpha\rightarrow\beta}\left(\varphi\right)$ in both TEAG-precursor and precursor-solid transitions during the critical period (Figure\ref{fig:figure2-glass}.b) in which glass \& crystal's non-liquid portion starts to deviate. Glass sample flattens while crystal sample keeps rising. This calculation is based on recorded excitations $\left\{\vec{r}\right\}_{t+1}-\left\{\vec{r}\right\}_{t}$ in simulations rather than imaginary ones $\sum A_{i}\left\{d\vec{r}\right\}_{i}$.

	Results clearly verify the above stated storyline. Energy barrier stops TEAG-precursor transition in glass (Figure\ref{fig:figure2-glass}.c). TEAG's presence also makes precursor-solid transitions less favorable in glass. (Figure\ref{fig:figure2-glass}.d) In all cases $dE\rightarrow0$ when $\varphi\rightarrow0$. As mentioned before this is reasonable since energy surplus compared to global minimal-energy configuration should approach zero when $\varphi=\partial G/\partial Q_{6}\rightarrow0$. Finally, it seems glassy dynamics is generally stronger when $\left|\varphi\right|$ is larger.
	
	Note that in Figure\ref{fig:figure2-glass}.d, $dE_{\alpha\rightarrow\beta}\left(\varphi\right)$ curve of glass is qualitatively different from that of crystal. This particular shape appears multiple times in different case studies. Its appearance is always linked to glassy systems in dynamical traps.
	
	\newpage
	\subsection{Supercooling \& Crystallization}
	Supercooling affects crystallization kinetics\cite{MooreMolinero2011}. Typical model is a probabilistic transition rate $k_{\alpha\rightarrow\beta}=\Delta N_{\alpha\rightarrow\beta}/N_{\alpha}dt \sim exp\left(-\Delta E_{\alpha\rightarrow\beta}/a\right)$. It's based on several layers of definitions, including morphology identifications and requirements for a valid transition event, among others. Here we apply our new thermodynamic framework for a clean picture. An abnormal sample is found during analysis, and its nature is instantly clear.
	
	Yukawa PMMA spherical colloidal samples at different volume fractions (supercooling degrees) are shear-melted. Their crystallization is observed under a Leica SP8 confocal microscope. Based on obtained coordinates, energy gaps as defined by equation \eqref{eq:6} are calculated for both liquid-precursor and precursor-solid transitions based on real recorded excitations. No distinction is made between BCC FCC and HCP. Volume fraction is signaled by cutoff radius $r_{c}$, as the first valley in radical distribution function $g\left(r\right)$.
	
	\begin{figure}[htbp]
		\centering
		\includegraphics[width=1\linewidth]{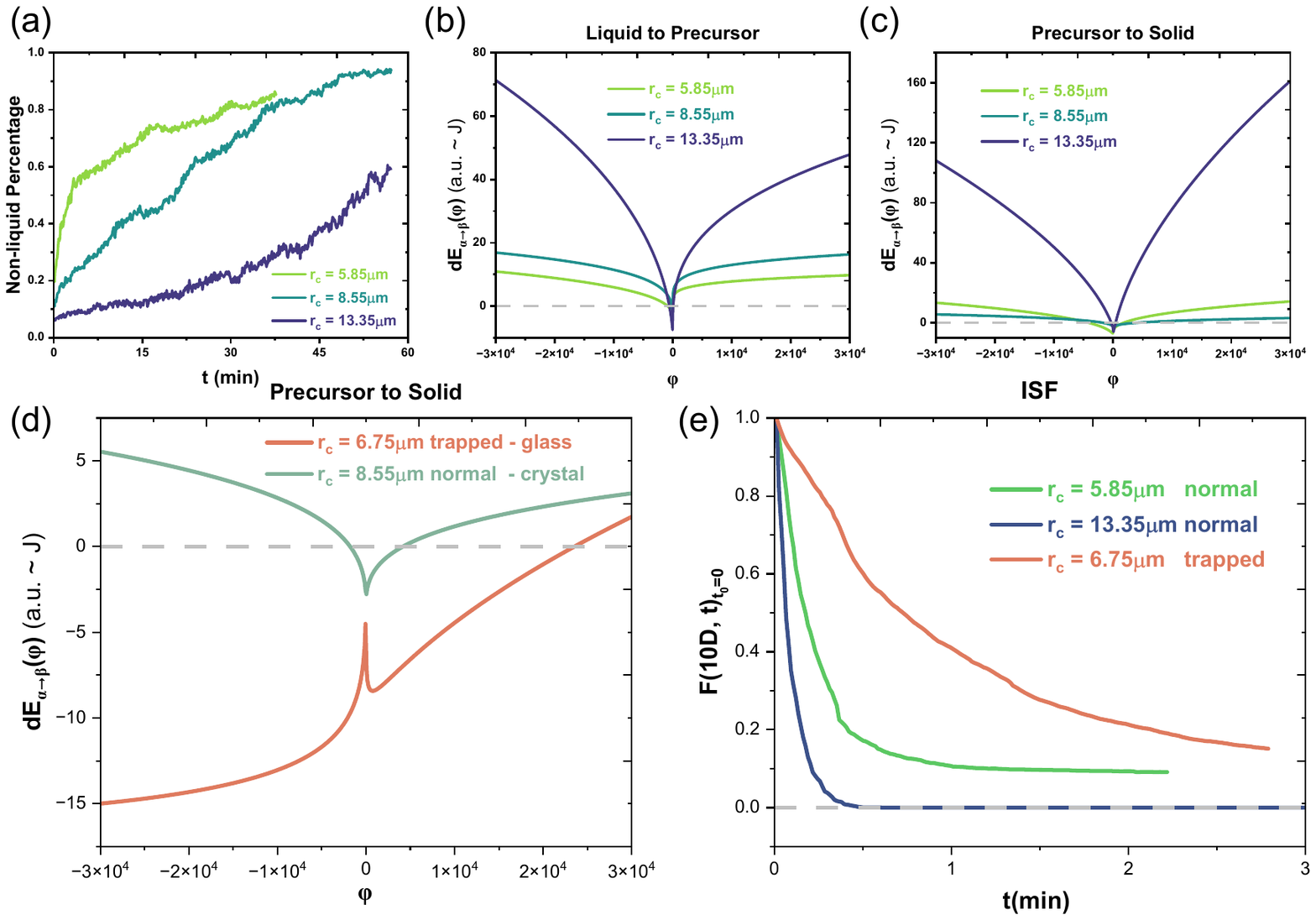}
		\caption{\textbf{Case study: supercooling.} Framework applications in crystallization kinetics. \textbf{(a) (b) \& (c) Supercooling accelerates crystallization.} As supercooling increases, signaled by increased volume fractions \& decreased $r_{c}$, non-liquid percentage grows more quickly (a). Energy gap for liquid-precursor (b) and precursor-solid (c) transitions is lowered, as expected. Morphology judge is based on BOO \& solid bond number. No distinction is made between BCC, FCC \& HCP. Calculations for energy gap is based on equation \eqref{eq:6} from real recorded excitations. \textbf{(d) \& (e) Yukawa glass-like slow relaxation.} An abnormal sample at $r_{c}=6.75\mu m$ is found with a qualitatively different $dE_{\alpha\rightarrow\beta}\left(\varphi\right)$ curve (d). This curve is similar to that of LJ glass in Figure\ref{fig:figure2-glass}.d. Thus we guess this is a Yukawa glass-like slow relaxation system. In (e), ISF calculation indeed confirms this hypothesis. Normal samples crystallize \& lose resemblance to liquid-dominated initial configurations. This process is faster in lower volume fraction $r_{c}=13.35\mu m$. Glass-like sample, on the other hand, is dynamically trapped whose resemblance lingers on.}
		\label{fig:figure3-volume-fraction}
	\end{figure}
	
	Reassuring results are shown in Figure\ref{fig:figure3-volume-fraction}.a \ref{fig:figure3-volume-fraction}.b \& \ref{fig:figure3-volume-fraction}.c. When supercooling degree increases, as indicated by increased volume fractions \& decreased $r_{c}$, crystallization kinetics accelerate: Non-liquid portion in samples grows more quickly. At the same time, $dE_{\alpha\rightarrow\beta}\left(\varphi\right)$ for both liquid-precursor \& precursor-solid transitions decreases, as expected, with the latter seems to be saturated \& roughly flattened from $r_{c}=8.55\mu m$ to $r_{c}=5.85\mu m$.
	
	An abnormal sample at $r_{c}=6.75\mu m$ is discovered. Its precursor-solid $dE_{\alpha\rightarrow\beta}\left(\varphi\right)$ curve is qualitatively different from other samples (Figure\ref{fig:figure3-volume-fraction}.d). Also, its non-liquid percentage grows slower than $r_{c}=8.55\mu m$ despite increased supercooling. Crystallization should accelerates, but the opposite is true.
	
	Under our framework, explanations are easily. The abnormal curve in Figure\ref{fig:figure3-volume-fraction}.d is similar to that of LJ glass in Figure\ref{fig:figure2-glass}.d. Thus we guess intuitively this sample must be dynamically trapped and forms a Yukawa glass-like slow relaxation system. This is verified by ISF $F\left(k=10D, t\right)_{t_{0}=0}$, as seen in Figure\ref{fig:figure3-volume-fraction}.e. Normal samples crystallize and lose resemblance to their liquid-dominated initial configurations at $t_{0}=0$. Lower volume fraction ($r_{c}=13.35\mu m$) speeds up this process because of elevated mobility, but generally ISF stabilizes within 30 seconds to 1 minute. However, the abnormal $r_{c}=6.75\mu m$ sample shows clear trapping behaviors. Its ultra-slow relaxation process continues at 3 minutes. Although this sample crystallizes eventually, it's abnormally slow dynamics did not escape our notice under $dE_{\alpha\rightarrow\beta}\left(\varphi\right)$ energy gap analysis. More physics remains to be unearthed as to the connection between glass-like dynamics and this particular shape of $dE_{\alpha\rightarrow\beta}\left(\varphi\right)$ (Figure\ref{fig:figure2-glass}.d \& \ref{fig:figure3-volume-fraction}.d). It could well be the first layer into an interesting field.
	
	Note the sign difference of energy gaps in LJ system (Figure\ref{fig:figure2-glass} - negative) and Yukawa system (Figure\ref{fig:figure3-volume-fraction} - positive). This may be caused by LJ force being repulsive and Yukawa force being attractive at short distances. This present another unsolved question. Be aware that $V\left(r\right)$ is \textit{not} directly included in energy gap calculation schemes of these two case studies. We only look at recorded movements. Only in demo calculations (Figure\ref{fig:figure1-demo}) is $V\left(r\right)$ used to calculate dynamical matrix \& imaginary excitations.
	
	\newpage
	\subsection{Curvature \& Crystallization}
	Confinement is a major topic\cite{SolanoCabrera2025Colloids}. Fundamental problems include how boundary influences bulk dynamics and what intriguing properties does the surface layer possess. Over the years, 3D \& 2D \& quasi-2D systems have been extensively studied in both hard (wall) and soft (electric field) confinement scenarios. Here we merely tap into this vast field by applying our framework to a simple case study. We investigate tube curvature's influence on crystallization dynamics and, if any, try to explain it with our framework.
	
	In a similar setup, PMMA Yukawa spherical colloid samples with $r_{c}=5.25\mu m$ and $D=2.5\mu m$ are sealed in circular tubes ranging from $d=100\mu m$ to $d=300\mu m$. Their crystallization dynamics is scanned under a Leica SP8 confocal microscope after shear melting by a $50\mu m$ nickel rod. FOV spans 7-8 layers from boundary into the bulk. Energy gap $dE_{\alpha\rightarrow\beta}\left(\varphi\right)$ for liquid-precursor transitions is calculated by the same scheme under equation \eqref{eq:6} using real recorded excitations. 
	
	\begin{figure}[htbp]
		\centering
		\includegraphics[width=1\linewidth, trim=0cm 3cm 0cm 5cm, clip]{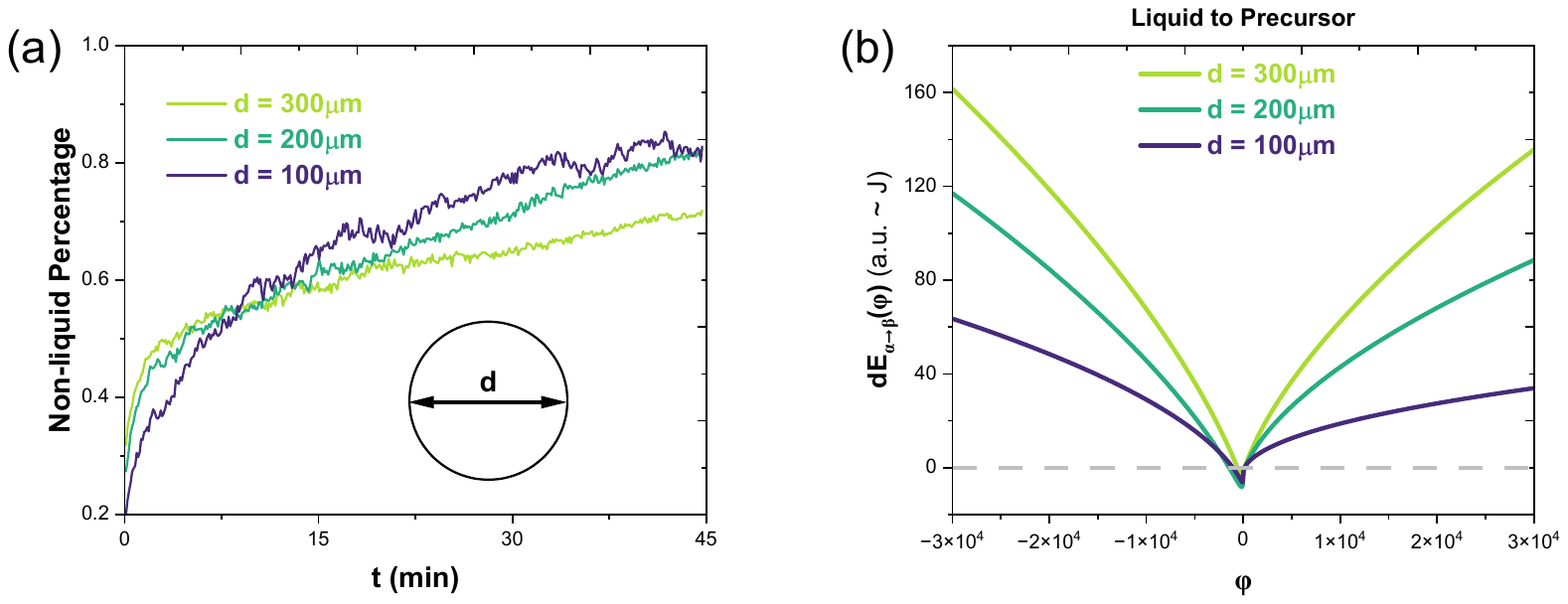}
		\caption{\textbf{Case study: curvature.} Framework applications in crystallization dynamics. Acceleration is clearly visible under tighter confinement or higher curvature in \textbf{(a)}. Smaller tubes are shear melted more thoroughly. They start from a lower non-liquid portion but still overtake later on. This phenomenon is clearly explained in \textbf{(b)}. Higher curvature corresponds to lower energy gap for liquid-precursor transitions. Calculations are based on experimental real recorded excitations.}
		\label{fig:figure4-curvature}
	\end{figure}
	
	As seen in figure\ref{fig:figure4-curvature}.a, higher curvature (lower $d$) accelerates crystallization. Smaller tubes are easier to shear melt thoroughly. They consequently start from a lower non-liquid portion but still overtakes larger tubes later on nonetheless. We are pleased to see this trend explained in energy gap calculations as in figure\ref{fig:figure4-curvature}.b. This serves as another fruitful application of our framework. Additionally, mobility is not involved here because ISF is identical for all samples.
	
	\section{Discussion}
	For more complicated topics, plausible variations of the above minimalist framework include $\left(1\right)$ order parameter $\alpha$ designed for a particular purpose, $\left(2\right)$ higher order terms of excess entropy, $\left(3\right)$ additional terms in equation\eqref{eq:1} for active matters or external fields, and $\left(4\right)$ a more sophisticated way of imaginary excitations generation. Hopefully, more insightful research can be built upon this work.

\end{document}